\documentclass[12pt]{article}
\usepackage{graphicx}
\usepackage{amsmath}   
\usepackage{url}
\usepackage{multirow}
\usepackage[colorlinks=true,linkcolor={black},urlcolor={black}]{hyperref}
\usepackage{ftnright}
\newcommand{\be}{\begin{equation}}
\newcommand{\ee}{\end{equation}}
\begin{document}

\title{Quantifying the Availability of TV White Spaces for Cognitive Radio Operation in the UK}

\date{\today}
\author{Maziar Nekovee \\ 
BT Research, Polaris 134, Adastral Park,\\
Martlesham, Suffolk, IP5 3RE, UK\\
and
\\
Centre for Computational Science, University College London\\
20 Gordon Street, London WC1H 0AJ, UK \\ 
maziar.nekovee@bt.com}
\maketitle

\begin{abstract}
Cognitive radio is being intensively researched for opportunistic access to  
the so-called TV White Spaces (TVWS): large portions of the  VHF/UHF TV
bands which become available on a geographical basis after the digital
switchover. 
Using accurate digital TV (DTV) coverage maps together with a  database
of DTV transmitters, we develop a methodology for
identifying TVWS  frequencies at any  given location in the United Kingdom.
We use our methodology to investigate variations in TVWS as a function of
the location and transmit power of cognitive radios, and examine 
how constraints on adjacent channel interference imposed by regulators 
may affect the results. 
Our analysis provides a realistic view on the  spectrum opportunity
associated with cognitive devices, 
and presents the first quantitative study of the
availability and frequency composition of TWVS outside the United States.

\end{abstract}

\section{Introduction}
Cognitive radio (CR) technology \cite{mitola1,haykin}
is a key enabler for the opportunistic spectrum access (OSA) model
\cite{horn,maz-tv}, a potentially revolutionary new paradigm for 
dynamic sharing of licenced spectrum with unlicensed
devices. In this operational mode a
cognitive radio acts as a spectrum scavenger. It performs
spectrum sensing over a 
range of frequency bands, dynamically identifies unused
spectrum, and then operates in this
spectrum at times and/or locations  when/where it is not used by incumbent
radio systems. Opportunistic
spectrum access can take place both on a temporal and a spatial
basis. In temporal opportunistic access a cognitive radio
monitors the activity of the licensee in a given location and uses the
licensed frequency at times that it is idle. An example of this is the
operation of cognitive radio in the radar and UMTS bands.
In spatial opportunistic access cognitive devices identify geographical regions  
where certain licensed bands are unused and access these bands without causing harmful
interference to the operation of the incumbent in nearby regions. 

Currently Cognitive radio is  being intensively researched for
opportunistic access to  the so-called TV White Spaces (TVWS): large portions of the  VHF/UHF TV
bands which become available on a geographical basis after the digital
switchover.
In the US the FCC (Federal Communications 
Commission) proposed to allow
opportunistic access to TV bands  already in 2004
\cite{fcc-cr}. Prototype cognitive radios operating in this mode were 
put forward to FCC by Adaptrum, I$^2$R, Microsoft, Motorola and Philips
in 2008 \cite{fcc-tvreport}.  After  extensive tests the FCC
adopted in November 2008 a Second Report and Order that establishes
rules to allow the operation of cognitive devices in TVWS on a
secondary basis \cite{fcc-tvdec}.
Furthermore, in what is potentially a radical shift in policy, 
in its recently released Digital Dividend Review Statement 
\cite{ofcom-ddr} the UK regulator, Ofcom, is proposing to ``allow licence exempt use of interleaved spectrum for cognitive
devices.'' \cite{ofcom-ddr}. Furthermore Ofcom states that 
``We see significant scope for cognitive equipments using interleaved 
spectrum to emerge and to benefit from international economics of
scale'' \cite{ofcom-ddr}. More recently,
on February 16 2009, Ofcom published a new consultation 
providing further  details of its proposed cognitive access to 
TV White Spaces \cite{ofcom-cr}.

With both the US and UK adapting the OSA model, and 
the emerging 802.22 standard for cognitive access to TV bands
\cite{ran1,ran3} being at the final stage, 
we can expect that, if successful, this new paradigm will become
mainstream among spectrum regulators worldwide.  However,
while a number of recent papers have examined various 
aspects of cognitive radio access to TVWS in the United States
\cite{new-america, fred, tv-alex,Siker,ofdm,motorola1}, there is
currently very little quantitative information  on the {\it global} spectrum
opportunities that may result if CR operation in TV bands becomes
acceptable in other countries in the world.

To bridge this gap, we present in this paper a 
quantitative analysis of TV White Spaces availability 
for cognitive radio access in the United Kingdom. 
Using 
accurate digital TV (DTV) coverage maps together with a  database
of DTV transmitters, we develop a methodology for
identifying TVWS  frequencies at any  location in the UK.
We use our methodology to investigate the variations in TVWS as a function of
the location and transmit power of cognitive radios, and examine 
how constraints on adjacent channel emissions of cognitive radios may 
affects the results. 
Our analysis provides a realistic view on the potential spectrum opportunity
associated with cognitive radio access to TWVS in the UK, 
and presents the first quantitative study of the
availability and frequency composition of TWVS outside the United States.

The rest of this paper is organised as follows. In
section II we discuss  in detail the operation of cognitive radio devices
in the VHF/UHF TV bands. This is followed by a description of our methodology for
estimating TWVS frequencies. Section III presents results of our
study of the availability of TVWS in $18$ locations in the UK and
analyses the implications of our findings. We conclude this
paper in Section V.

\section{ Cognitive radio operation in TV bands}
Broadcast television services operate in licensed channels in the VHF and UHF portions of the radio spectrum. The regulatory 
rules in most countries prohibit the use of unlicensed devices in TV
bands, with the exception of remote control, medical telemetry devices
and wireless microphones. In most developed countries regulators are currently in the
process of requiring TV stations to convert from analog to digital
transmission. This {\it Digital Switchover} is  expected to
be completed in the US in 2009 and in the UK in 2012. A similar
switchover process is also underway or being planned (or is already completed)
in the rest of the EU and many other countries around the world.

After Digital Switchover a portion  of TV analogue channels
become  entirely vacant  due to the higher spectrum efficiency of digital TV (DTV). These
cleared channels will then be reallocated by regulators to other services,
for example through auctions. In addition, after the DTV
transition there will be typically a number of TV channels in a  given geographic area that are not being used by DTV
stations, because such stations would not be able to operate without
causing interference to co-channel or adjacent channel stations. These
requirements are based on the assumption that stations operate
at maximum power. However, a transmitter operating on a vacant TV
channel at a much lower power level would not need a great separation
from co-channel and adjacent channel TV stations to avoid causing
interference. Low power unlicensed  devices can operate on vacant channels in
locations that could not be used by TV stations due to interference
concerns \cite{new-america}.
These vacant TV channels are known as TV White Spaces, or interleaved
spectrum in the parlance of the UK regulator.
Opportunistic operation of  cognitive
radios in TV bands, however, is conditioned on the ability of these
devices to avoid harmful interference to licensed users of these
bands, which in addition to DTV include also wireless microphones
\cite{ new-america}. 
In November 2008, the FCC adopted a report setting out rules allowing licence-exempt cognitive 
devices to operate in TV White Spaces. In summary these rules require cognitive devices to 
use  a combination of spectrum sensing and geolocation. The devices must be able to sense both TV signals and wireless 
microphones down to $-114$ dBm, and must also locate their position to within $50$ meters accuracy and then consult a database that 
will inform them about available spectrum in that location. Devices without geolocation capabilities are also allowed if they are 
transmitting to a device that has determined its location. Cognitive devices that use sensing alone are allowed in 
principle. However, the FCC states that such devices will be ``subject to a much more rigorous approval process'' \cite{fcc-tvdec}.

The fundamental reason why TVWS have  attracted much interest is
an exceptionally attractive combination of bandwidth and
coverage. Signals in TV bands, travel much further than both the WiFi and 
3G signals and penetrate buildings more readily. This in turn means
that these bands can be used for a very wide range of potential new
services, including last mile wireless broadband in urban environments, broadband wireless
access in rural areas \cite{ran1,ran3}, new types of mobile
broadband and wireless networks for digital homes. Furthermore, in the case of the
UHF bands, the wavelength of signals in these bands is sufficiently
short such that resonant antennas with sufficiently small footprint can be
used which are  acceptable for many portable use cases and handheld
devices \cite{fred}.

\begin{figure}
\centering
\begin{tabular}{@{}ccl@{}}
\includegraphics[width=5.6in]{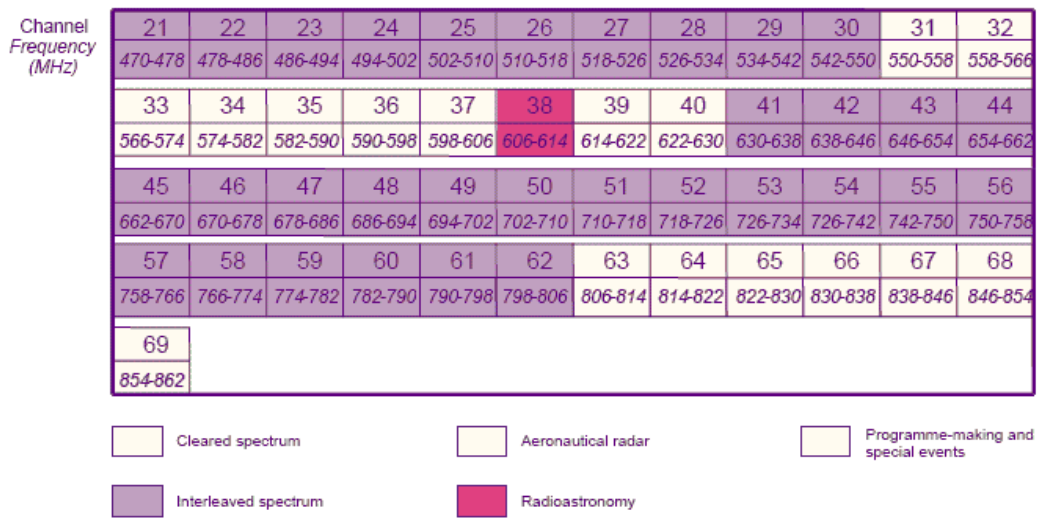} 
\end{tabular}
\caption{The available VHF and UHF TV spectrum in the UK after digital switchover, 
showing both the interleaved (TVWS) and cleared channels \cite{ofcom-ddr}.}
\label{fig:specpie}
\end{figure}

\section{Methodology}
The digital TV standard adopted in the UK and the rest of Europe is
DVB-T (Digital Video Broadcasting Terrestrial) which uses $8$ MHz wide
frequency bands for its transmission. 
This is unlike the US ATSC standard where each band is $6$ MHz wide.
Fig. 1 shows the chart of the UK's analog TV frequency bands and how these will be
divided after digital switchover into cleared and interleaved spectrum
\cite{ofcom-ddr}. From this chart it can be seen that the
total UK interleaved spectrum, which is entirely in the UHF 
frequency range, is $256$ MHz. However, Ofcom has proposed to auction
off channels $61$ and $62$ for licenced use \cite{ofcom-interleaved}, reducing 
the TV bandwidth available for access by cognitive devices to  a total of
$240$ MHz. However the exact number and frequency composition of TVWS
can vary from location to location and is 
determined by the spatial arrangement of DTV transmitters and their 
nationwide frequency planning. 

The CR transmission at a given location should not cause harmful interference to TV receivers
both within the coverage area of nearby transmitters, and at the edge 
of this area. To achieve this the CR device can
transmit on the TV bands used by these transmitters only if its position
is a  minimum ``keep-out'' distance, $R_{cr}$, away from the  edge of
their coverage area \cite{ran3}.
Fig. 2 shows schematically a typical setup for the operation of 
a cognitive radio base station which operates in a given location 
in TV White Spaces which are available at that location.

\begin{figure}
\centering
\begin{tabular}{@{}ccl@{}}
\includegraphics[width=5.6in]{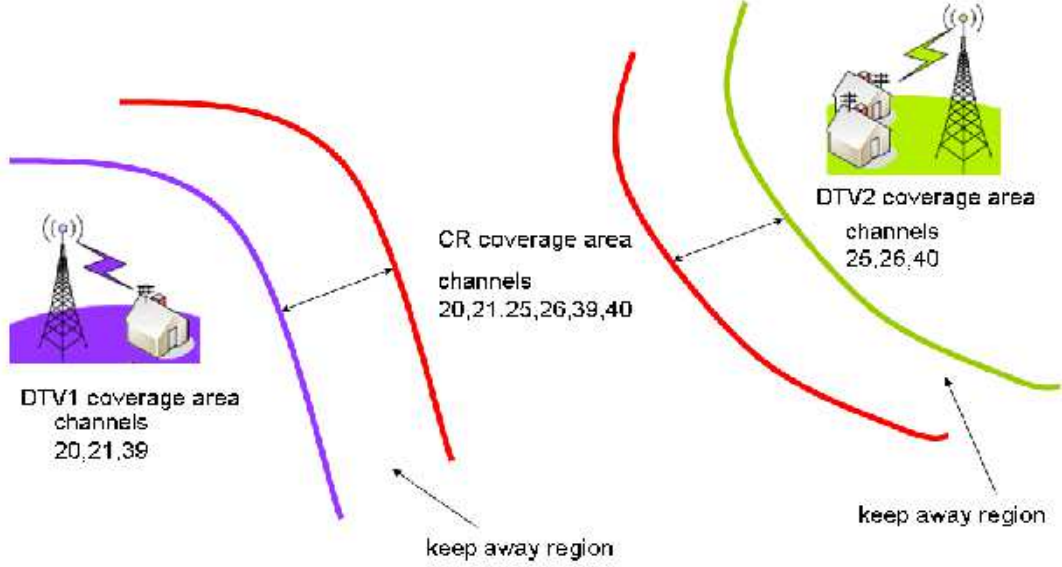} 
\end{tabular}
\caption{Opportunistic access to interleaved TV spectrum (White Spaces) 
by cognitive radios.}
\label{fig:specpie}
\end{figure}

In a simplified picture, based on the pathloss model \cite{pathloss}, the
keep-out distance can be obtained as follows.  Denote with $R_{tv}$ the
maximum coverage radius of the TV station, and with $P_{cr}$ and $P_{tv}$ the
transmit power of the TV transmitter and the CR transmitter,
respectively.  Then, in order to avoid interference with TV receivers that are 
at  the edge of the coverage area, we must have: 

\begin{equation}
\frac{ P_{\text{tv}} /R_{\text{tv}}^\alpha }
      {P_{\text{cr}}/R_{\text{cr}}^\alpha }\geq \beta_{\text{th}},
\end{equation}
where $\beta_{th}$ is the sensitivity threshold of a TV
receiver, and $\alpha$ is the pathloss exponent. This yields: 
\begin{equation}
R_{cr}\geq \left 
( \beta_{\text{th}} \frac{P_{\text{cr}}}{P_{\text{tv}}}\right)^{1/\alpha} 
R_{\text{tv}}.
\end{equation}
Consequently, a CR device at location ${\bf r}$ can use the frequencies
associated with a TV station located at ${\bf R}_j$ only if 
$|{\bf r}-{\bf R}_j| \geq  R'_j$, where 
\begin{equation}
R'_j= \left[ 1+
\left ( \beta_{\text{th}} \frac{P_{\text{cr}}}{P_{
\text{tv}}^j}\right)^{1/\alpha}\right] R_{\text{tv}}^j.  
\end{equation}
Repeating the above procedure for every TV transmitter, 
one can obtain the total number of TV transmitters
whose associated frequencies can be used by a CR
operating with a specified transmit power, $P_{cr}$, at location ${\bf r}$, 
from which the total number of TVWS
frequencies  available for opportunistic access, $\rho({\bf r},P_{cr})$, can be obtained as:
\begin{equation}
\rho({\bf r},P_{\text{cr}})= 
\sum_j \sum_m \Theta(|{\bf r}-{\bf R_j}|-R'_j) \delta_{mj}
\end{equation}
where $\Theta$ is the step function and $\delta_{mj}=1$ if a frequency
$f_m$ is used  by a DTV transmitter located at $R_j$ and zero otherwise.
Furthermore
the first and the second sum in the above equation are  over all DTV
transmitters and all DTV frequencies, respectively.

In reality coverage contours of TV transmitters are far from
circular due to a combination of terrain and clutter (building, trees, etc) 
diffraction
of radio waves, non-isotropic radiation patterns of transmitter antennas, 
and interference resulting from nearby DTV transmitters
\cite{bbc}. 
Furthermore, shadow fading and atmospheric effects give rise to
stochastic fluctuations in the received TV signal power \cite{bbc}.
In our study we make use of the publicly available maps of DTV
coverage in the UK \cite{freeview}
which  were generated via computer simulations from the
Ofcom's database of location, transmit power (ERP), antenna height
and transmit frequency of UK's $81$ main DTV  transmitters. These 
computer simulated coverage maps are further validated  and refined
through measurements and direct observation by DTV users. Fig. 3 shows, as an 
example, the coverage map of a  DTV transmitter located in the vicinity of 
Oxford \cite{freeview}.

The typical transmit power of UK's DTV transmitters ranges 
between $25$W and $200$ kW (ERP) \cite{ofcom-trans}.
Consequently in the case of cognitive devices with 
transmit powers typical of licenced-exempt usage 
($\sim 100$ mW) we  have $P_{CR}/P_{TV} \ll 1$. Consequently, 
in this low-power limit an {\it upper bound for} 
the vacant TWVS frequencies at a given  location can be directly extracted 
from the coverage maps of DTV transmitters, as can also be seen from 
Eqs.(3-4).

Our computer algorithm for obtaining such upper bounds works as
follows. We use the UK National Grid (NG) coordinate 
system \cite{ngc} in order to specify the geographical 
position of any location on the UK map. 
Given the NG coordinates of a UK location our
code then maps this location onto the closest grid point on the coverage
maps of DTV transmitters.  For a given DTV transmitter
this grid point is then evaluated to
determine if it  falls within the coverage area of that
transmitter. If this is the case, then the frequencies associated 
with the transmitter are tagged as {\it occupied}  
at that location, otherwise they are tagged as {\it vacant}. 
Repeating this procedure for coverage maps of
the entire $81$ UK transmitters, we then obtain 
a list of vacant TV frequencies at a given location 
that can be used by a low-power cognitive device which is positioned 
in that location.

\begin{figure}
\centering
\begin{tabular}{@{}ccl@{}}
\includegraphics[width=4.2in]{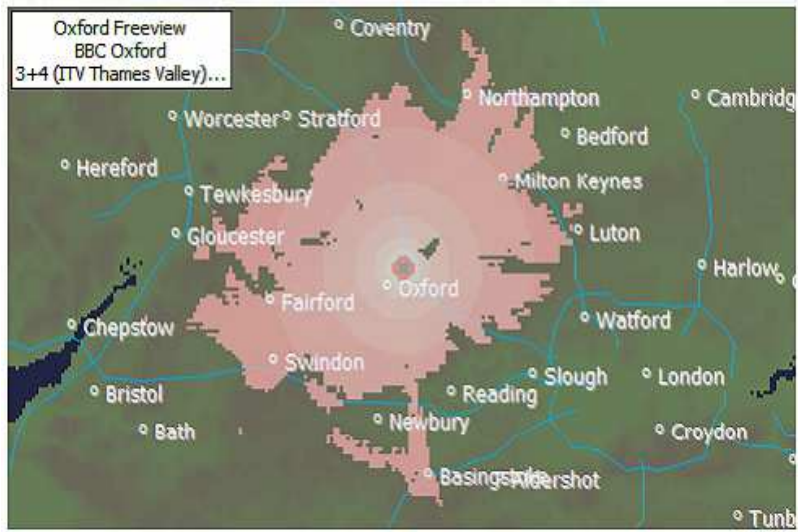} 
\end{tabular}
\caption{
Coverage map of the DTV transmitter located  near Oxford 
is shown. The square marks the location of
the transmitter. The coverage area is shown in pink \cite{freeview}.}
\label{fig:cp}
\end{figure}
In the case of high power cognitive equipments, e.g. those considered
within the 802.22 standard  with transmit powers as high as 
$4$ Watts (ERP) \cite{ran3}, 
the keep-out radius associated with a cognitive radio
is substantial, e.g. between $30-120$ km.
This keep out radius need to be taken into account when
estimating vacant TV frequencies at a given location. However, due
to the irregular shape of TV coverage contours the required
computations are  very intensive. In order to reduce
this computational effort, we have approximated the actual DTV coverage areas
by circular disks centred at a each DTV transmitter.
These disk were  constructed such that each of them 
entirely encompassed the coverage area of the associated transmitter
while also having the minimum possible surface area. 
With this simplification, it is then computationally straightforward to 
calculate from Eqs. (3-4) the vacant TV frequencies as a function of 
both position and transmit power of cognitive devices.

\section{Results}
Potential applications of TVWS devices  
will strongly depend on how the availability of 
this spectrum varies, both  from location to
location and  as a function of transmit power of cognitive devices.
The FCC, for example, 
is considering two classes of uses cases. The first corresponds
to fixed devices with relatively high transmit power, line of sight 
operation and ranges up to $30$ km. One expected use case for this
class is broadband wireless access to rural areas, for which the IEEE
802.22 standard is being developed. A second class of use cases under
consideration by the FCC are those associated with personal/portable
devices which maybe nomadic or mobile \cite{fred}.
In the rest of this paper we shall focus mainly on use cases corresponding 
to low-power cognitive access. A full  analysis for the case of 
high transmit power devices will be presented elsewhere.

\subsection{TVWS availability and frequency composition}
Fig. 4 summarises in a bar-chart  the availability of 
TVWS channels for $18$ major population centres in England, Wales and 
Scotland. The total number of channels available at each location is  
shown as red bar. It can be seen that there are considerable 
variations in the number of TVWS channels as we move from one UK location 
to another. For any given location, however, a minimum of $12$ channels
($96$ MHz) are accessible to low-power cognitive devices, while the average 
available spectrum is just over $150$ MHz.

\begin{figure}[h]
 \includegraphics[width=4.0in,angle=-90]{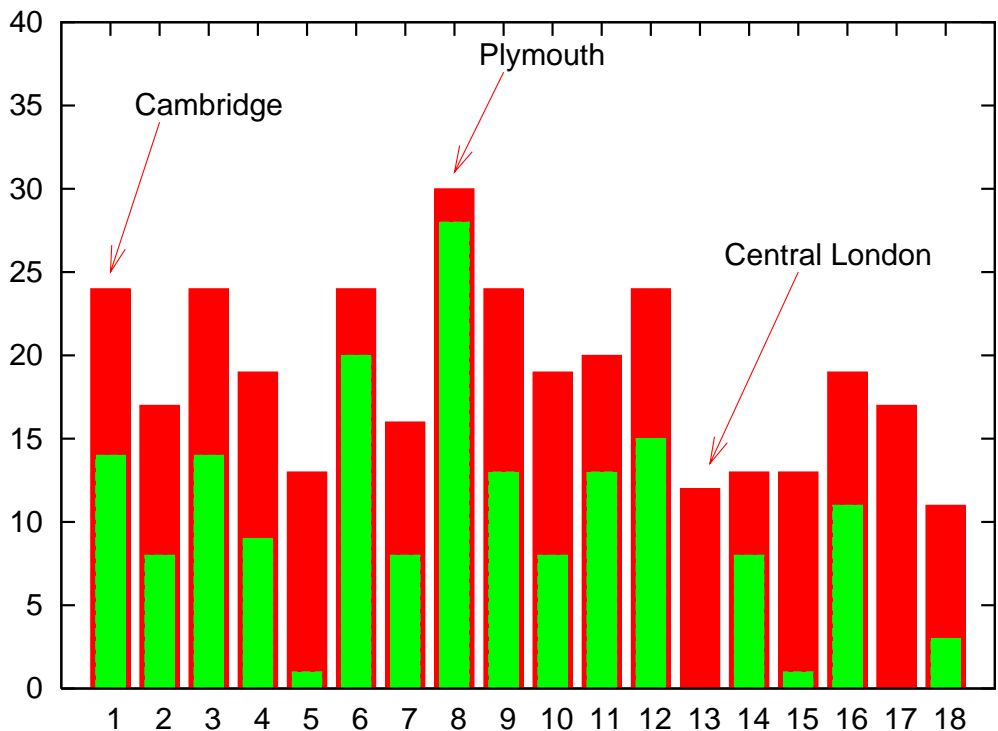}
   
 \caption{Variations in the availability of TV White Spaces is shown for 18 UK locations. Results are 
 shown before (red bars) and after (green bars) the exclusion of those vacant channels whose adjacent channels were 
 found to be occupied by DTV transmission.}
\end{figure}

In addition to estimating total available TVWS, it is of importance 
to investigate channel composition of this spectrum. 
In Fig. 5 we show, as an example, channel composition of TVWS 
in $4$ cities in England: Bristol, Liverpool, London and
Southampton. In this Figure vacant channels are shown as blue bars 
while occupied channels are left blank. As can be seen from 
the figure, the precise composition of TVWS channels vary greatly from location 
to location. In particular, both in  Bristol and Liverpool most
of the available channels are located in the lower end ($470-550$ MHz)
of the UHF band while in the case of Southampton these channels are bunched up 
in the higher end of this band ($630-806$ MHz).
Furthermore, the available TVWS channels can be highly {\it non-contiguous}. This feature may greatly restrict access to TVWS 
by most current wireless technologies, as modulation schemes implemented 
in these technologies often require a contiguous portion of the spectrum.
In the case of London, for example, although a total of 
$96$ MHz spectrum is in principle available, only $16$ MHz can be utilised for contiguous frequency access.
\begin{figure}[h]
 	 \begin{tabular}{@{}ccl@{}}
\includegraphics[width=1.7in,angle=-90 ]{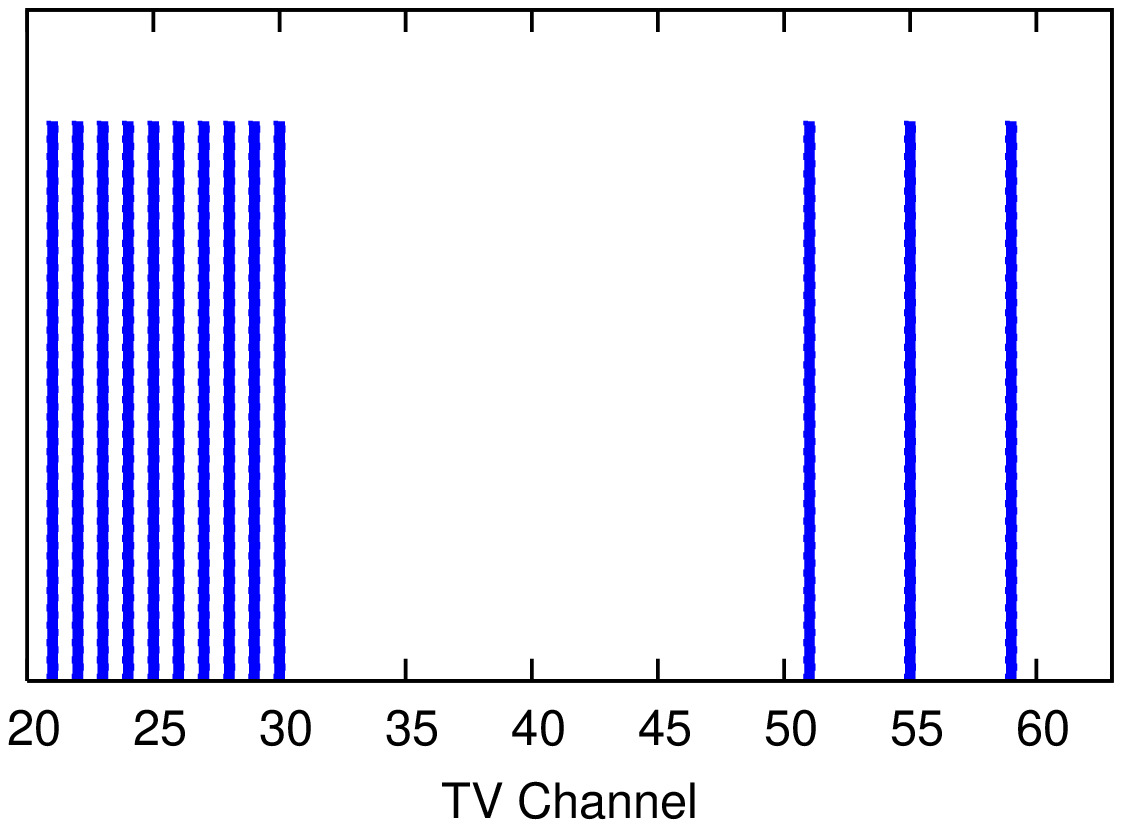} & 
\includegraphics[width=1.7in,angle=-90]{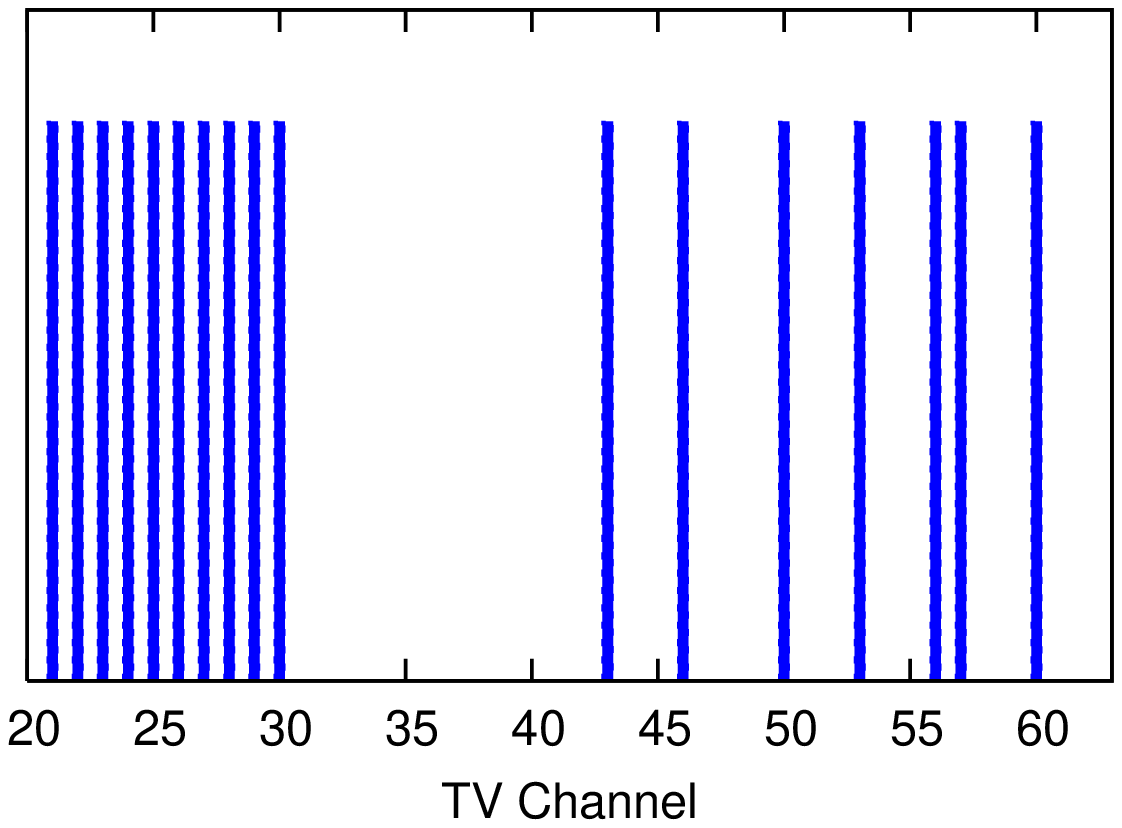}\\
\includegraphics[width=1.7in,angle=-90]{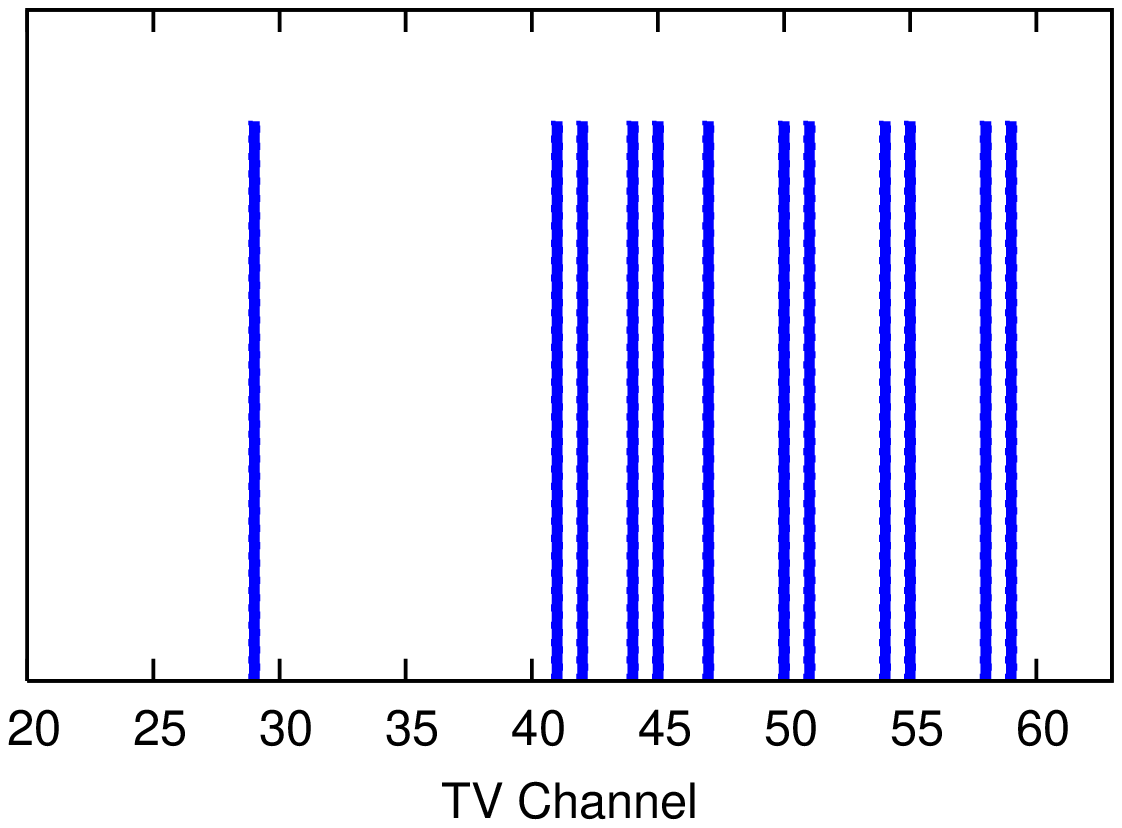}&
\includegraphics[width=1.7in,angle=-90]{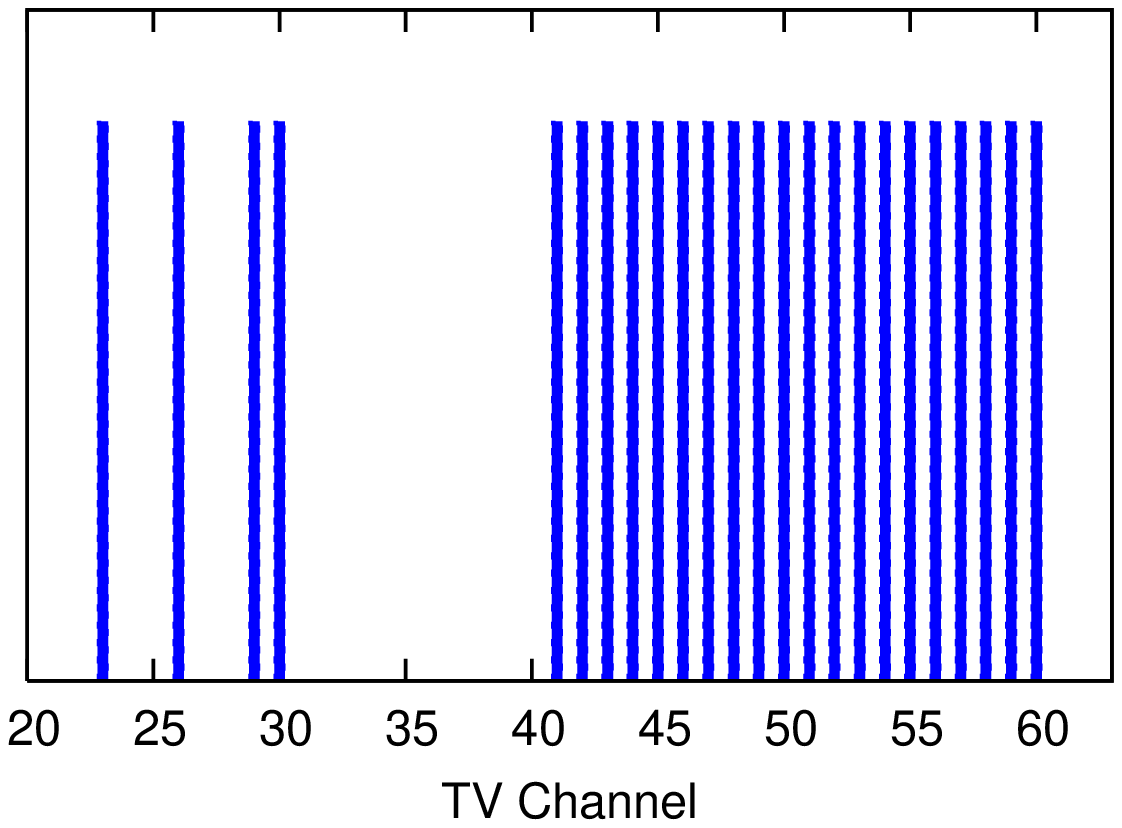}  
  
   	\end{tabular}
   \vfill
 \caption{The availability of TV White Space frequencies for low power
 cognitive radios are shown, from left to right and top to bottom, in
 Bristol, London, Liverpool and Southampton. Channels available for cognitive radio are shown as blue bars.} 
\end{figure}

\subsection{The impact of adjacent channel interference and transmit power}
When a high power cognitive device operates in a vacant TV
channel, energy leakage to adjacent channels may cause interference to TV 
sets that are tuned to these adjacent frequencies. To eliminate the occurrence of 
such adjacent-channel interference  the IEEE 802.22 Working Group prescribes that 
if channel $N$ is occupied by an incumbent, then cognitive devices should not only vacate this  channel  
but they also should refrain from transmitting at 
channels $N\pm1$. In addition, in the UK Ofcom has raised concerns 
that operation of low-power cognitive devices on a given channel may also cause adjacent-channel
interference for mobile TV receivers that are in close vicinity. 
Consequently, even in such 
low-power use cases, cognitive devices may be be constraint not to use vacant channels whose
immediate adjacent frequencies are used for mobile TV.
It is therefore of considerable interest to investigate how imposing such 
constraints will affect the availability of TVWS spectrum.

The  total number of available TVWS after imposing the above adjacent channel constraint are 
shown as green bars in Fig. 6. These results are obtained by eliminating from the list of TVWS channels 
any vacant channel whose immediate adjacent channels were found occupied.
It can be seen that imposing the constraint drastically  reduces 
the amount of accessible  spectrum in most locations considered.
In particular, in the case of central London we see that with this 
constraint imposed there will be {\it no channel} available for the operation of CR devices.
Averaging the results over all locations, we find that with this constraint 
imposed there will be only  $\sim 30$ MHz of TV spectrum available for cognitive access. 

Finally, we use the estimation approach outlined in Section  
III to briefly examine the impact of cognitive radio transmit power on the 
the availability of TVWS channels. Fig. 7 shows, as an example, variations 
in the number of vacant TV channels as a function of CR transmit power 
in the case of Manchester \cite{yang}. 
It can be seen that for $p_{cr} \leq 100$ mW
up to $17$ channels (136 MHz) are available. However, the availability 
decreases sharply as the transmit power is increased beyond this range.
Nevertheless, there is still $~ 40$ MHz of spectrum available for 
CR device transmitting  at $2$ W, within the typical operation of future 802.22 devices.

\begin{figure}
\centering
\begin{tabular}{@{}ccl@{}}
\includegraphics[width=4.2in]{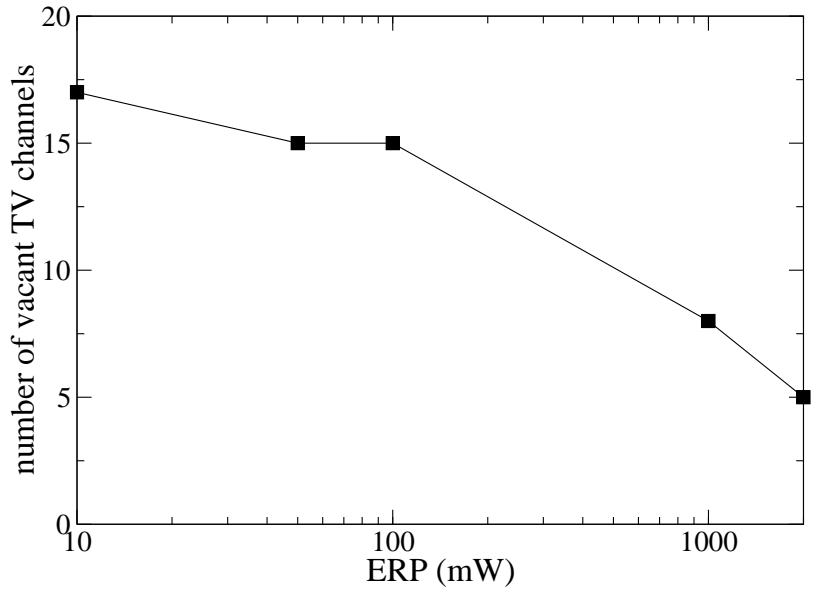} 

\end{tabular}
\caption{
Variation in the number of TVWS channels is shown  as a function of 
transmit power, computed for the city of Manchester.}
\end{figure}

\section{Conclusions}
In this paper we presented a methodology for estimating the UK TV White Spaces
for opportunistic access by cognitive radios. 
Using our methodology we examined the availability of  this
spectrum and its channel composition in $18$ UK population centres.
Our analysis shows on  average $\sim 150$ MHz of TVWS is 
available for access 
by low-power cognitive radios.
We found, however, that in many locations this considerable bandwidth is 
fragmented into many non-adjacent channels. Consequently, we conclude 
that the availability of novel pooling techniques, such as NC-OFDM
\cite{ofdm,nc-ofdm} is crucial for effective utilisation of TVWS, 
in particular for future high bandwidth applications. Finally, we examined 
the effect of constraints on adjacent channel interference imposed by 
regulators/standards on TVWS, and showed that such constraints 
drastically  reduce the availability of this spectrum.

Most future use scenarios of TVWS  will involve multiple cognitive devices
operating within the same geographical region \cite{sai}.
Some of our future work will focus on new methodologies for estimation 
and control of {\it aggregated} cognitive radio interference in such 
scenarios. We are also working on improved TVWS estimation methods  for high power use cases.

\section*{Acknowledgements}
The author is grateful to Brian Butterworh (UK Free.TV) for  
providing access to his DTV coverage data, and to 
Keith Briggs (BT), Yang Fang and Anjum Pervez (South Bank University) 
for their valuable contributions.

\end{document}